\documentclass[twoside]{dis09}
\usepackage[latin1]{inputenc}
\usepackage[dvips]{graphicx,epsfig,color}
\usepackage{wrapfig,rotating}
\usepackage{amssymb,amsmath,array}

\begin{document}

\title{The $Z$ boson $a_T$ distribution at hadron colliders}

\author{Andrea Banfi $^1$, Mrinal Dasgupta and Rosa Maria Duran Delgado $^2$
%
%
\vspace{.3cm}\\
%
1- Universit\`a di Milano-Bicocca and INFN, Sezione di Milano-Bicocca, Italy \\
Piazza della Scienza 3, 20126 Milano, Italy.
%
\vspace{.1cm}\\
2- School of Physics and Astronomy, University of Manchester\\
Manchester Oxford road, M13 9PL, United Kingdom.\\
}

\maketitle

\begin{abstract}
  We provide a theoretical study of a novel variable introduced in
  Ref.~\cite{WV} to study the transverse momentum of the Z boson at
  hadron colliders. The variable we consider has experimental
  advantages over the standard $p_T$ distribution enabling more
  accurate measurement at low $p_T$. We provide an all orders
  perturbative estimate for this variable at the next-to--leading
  logarithmic accuracy and compare the results to those for the
  standard $p_T$ distribution. We test our resummation, at the
  two-loop level, by comparing its expansion to fixed-order
  perturbative estimates and find agreement with our expectations.

\end{abstract}

\section{Introduction}
The transverse momentum $p_T$ of lepton pairs (or equivalently that of
gauge bosons which decay to lepton pairs) produced in hadron-hadron
collisions has been one of the most actively studied variables in QCD
phenomenology over several decades. In spite of sustained activity in
this area it remains a subject of vital importance and interest. An
accurate understanding of the production rates and $p_T$ spectra of
the $W$ and $Z$ boson can play a role in diverse applications such as
luminosity monitoring at the LHC to the measurement of the $W$ mass
and most importantly perhaps in the search for new physics which may
manifest itself for instance in the decay of new gauge bosons to
lepton pairs.

Moreover such spectra have been a source of considerable information
on QCD dynamics. Of particular interest here is the low $p_T$ region
where the resummation of logarithmically enhanced terms reflects an
understanding of QCD to all orders in the soft/collinear limit. The
low $p_T$ region is also of interest in constraining the
non-perturbative ``intrinsic $k_t$'' which is relevant in many other
hadron collider observables. The $p_T$ spectrum is consequently also
of immense use in the tuning of Monte Carlo event generators both for
the shower parameters and the intrinsic $k_t$ component. It is
therefore clear that as accurate a measurement of the $p_T$ spectrum
as possible is desirable especially at low $p_T$. However
comparatively large experimental errors have somewhat impacted
conclusions about the low $p_T$ region, which have thus not been
optimal.

An observable that may offer experimental advantages over the $p_T$
variable has been suggested in Ref.~\cite{WV}. The variable in
question is simply the component of the $p_T$ distribution transverse
to an axis also defined using the lepton transverse momenta. This
component, called the $a_T$ can be measured more accurately than the
longitudinal component $a_L$ and hence than the overall $p_T
=\sqrt{a_T^2+a_L^2}$~\cite{WV}.  Consequently the phenomenology of the
$a_T$ variable assumes some importance.

In the current article we provide one of the most important
ingredients necessary for an accurate investigation of the $a_T$
variable over the full range of measured values, the resummation of
logarithms in $a_T$ up to the next-to--leading logarithmic (NLL)
accuracy. We shall in forthcoming work combine these estimates with
fixed-order results and include non-perturbative effects to obtain a
state of the art prediction. The resummation that we carry out
involves some important differences from the classic $p_T$ resummation
formalism~\cite{DDT,CSS}, which we explain.

\section{The $a_T$ distribution}
To calculate the $a_T$ distribution at low $a_T$ one first needs to
obtain the dependence of the observable on the momenta of soft emitted
gluons.  Denoting the transverse lepton momenta by $\vec p_{t1}$ and
$\vec p_{t2}$ we define the ``thrust'' axis as in Ref.~\cite{WV}
\begin{equation} 
\label{eq:axis} 
\hat{n} = \frac{\vec{p}_{t1}-\vec{p}_{t2}}{|\vec{p}_{t1}-\vec{p}_{t2}|}.
\end{equation} 
Considering the emission first of a soft gluon with transverse
momentum $k_t$, one obtains that the $a_T$ with respect to the above
defined axis is simply
\begin{equation}
a_T = k_t |\sin \phi| +\mathcal{O}\left(k_t^2\right),
\end{equation}
where at NLL accuracy we can neglect the order $k_t^2$ correction,
while the usual $p_T$ variable is just given by $p_T =k_t$ and $\phi$
is the angle between the soft gluon and the thrust axis in the plane
transverse to the beam. At all orders the above result generalises to
\begin{equation}
a_T = \left|\sum_i k_{ti}\sin \phi_i\right|= \left|\sum_i k_{xi}\right|\,.
\end{equation}

The most noteworthy aspect of the above result is that the $a_T$
variable depends on the sum over a {\emph{single}} component of soft
gluon momenta (along say the $x$ axis as indicated above). This is
different from the $p_T$ variable which at all orders is the
two-dimensional vector sum of individual gluon momenta $p_T =
|\sum_i{\vec{k}_{t,i}}|$.  The resummation of observables involving a
one-dimensional sum has been encountered before (see
e.g.~Refs.~\cite{BMS,BDET,BDD}) and we employ the formalism detailed
there to arrive at our result quoted in the following section.

\section{Resummed result}
Here we quote the NLL resummed result for the $a_T$ distribution.  As
for the $p_T$, it involves resummation in impact-parameter or $b$
space which is the Fourier conjugate of $k_t$ space and the resummed
result takes the form
\begin{equation}
\Sigma_N(a_T) = \sigma_0(N)\, W_N(a_T)\,, 
\end{equation}
where $\Sigma$ denotes the cross-section for events below some fixed
value $a_T$ (i.e the differential cross-section integrated over a
limited range from $0$ to $a_T$). For simplicity we have chosen to
express our result in moment ($N$) space conjugate to $\tau=M^2/s$,
with $M^2$ the mass of the lepton pair and where $\sigma_0$ is the
Born cross-section.  The function $W_N(a_T)$ contains the resummation
of large logarithms in $b$ space and reads
\begin{equation}
\label{eq:rad}
W_N(a_T) = \left(1+C_1(N)\, \bar{\alpha}_s\right)\, \frac{2}{\pi}\int_0^{\infty} \frac{db}{b} \sin(b a_T) e^{-R(b)}.
\end{equation}
\begin{wrapfigure}{r}{0.5\columnwidth}
\centerline{\includegraphics[width=0.45\columnwidth]{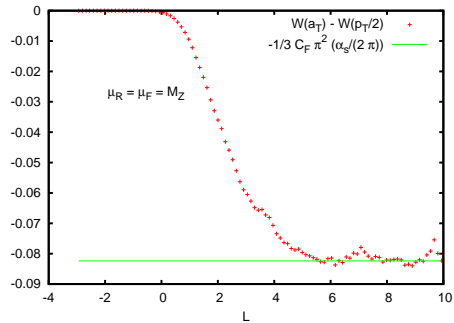}}
\caption{The difference between the integrated cross-sections for
  $a_T$ and $p_T/2$ at leading order in $\alpha_s$. The green line is
  our expectation for large positive $L$ and the red data points are
  the result from MCFM.}
\label{fig:lo}
\end{wrapfigure}

\begin{wrapfigure}{r}{0.5 \columnwidth}
\centerline{\includegraphics[width=0.45\columnwidth]{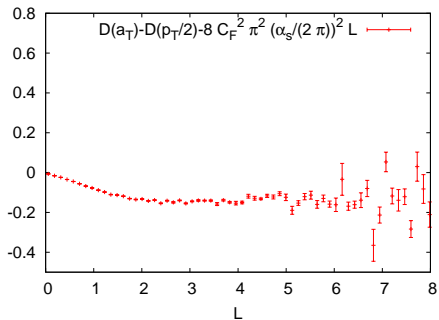}}
\caption{The difference between MCFM and our theoretical estimate for
  the difference in the distributions for $a_T$ and $p_T/2$. The
  result seems to be consistent with the expectation of an
  asymptotically flat distribution.}
\label{fig:nlo}
\end{wrapfigure}
The NLL $b$ space resummation gives rise to the exponential of $R(b)$,
the ``radiator'', precisely as in the standard $p_T$ resummation. The
difference with the $p_T$ distribution arises from the $\sin (b a_T)$
function that represents the conservation of a single component of
transeverse momentum. This is in contrast to the function $J_1(b p_T)$
that one encounters in $p_T$ resummation, which represents
conservation of both components of transverse momentum, relevant in
that case. Explicit computation reveals that, within NLL accuracy, the
radiator $R(b)$ in the above case turns out to be identical to that
for the $p_T$ variable so all differences are due to the presence of
the $\sin(b a_T)$ as opposed to a Bessel function. The term
$1+C_1 \bar{\alpha}_s$ represents a multiplicative correction which
can be obtained by carrying out a full leading-order calculation (here
$\bar \alpha_s =\alpha_s/2\pi$).

The main difference of the above result from the $p_T$ distribution
will be the absence of a Sudakov peak in the final result for the
$a_T$ differential distribution with the result continuously rising to
a constant value at $a_T=0$. The absence of the Sudakov peak is due to
the fact that the one-dimensional cancellation of a component of the
$\vec k_{ti}$ is the dominant mechanism for producing a low $a_T$
rather than Sudakov suppression. This is not true in the case of the
$p_T$ until very low values of $p_T$ beyond the Sudakov peak and hence
the formation of a peak in that case.

\section{Comparisons to fixed order estimates}
Here we compare our predictions to those from fixed order Monte Carlo
computations from the code MCFM \cite{EllisCamp}. To this end we
expand Eq.~\eqref{eq:rad} to order $\alpha_s^2$. After noting that the
result closely resembles the analogous result for the $p_T$
distribution with the substitution $a_T \to p_T/2$ it proves most
convenient to provide our predictions for the difference between the
distributions for $a_T$ and $p_T/2$.  Up to the two-loop level we
predict for this difference
\begin{equation}
\label{eq:pred}
W(a_T)-\tilde W\left(\frac{p_T}{2}\right)|_{\frac{p_T}{2}=a_T} =-\frac{1}{3}{\pi^2} C_F \bar{\alpha}_s+4\pi^2 C_F^2 \bar{\alpha}_s^2 L^2+\mathcal{O}(\alpha_s^2 L).
\end{equation}
In the above $\tilde W$ denotes the integrated event fraction for the
$p_T$ variable, which is widely available in the literature
($\bar{\alpha}_s =\alpha_s/2\pi$ and $L \equiv \ln M/a_T$).
 
We can test our resummation for the $a_T$ variable by comparing the
above result to the fixed-order computation from the program MCFM. At
the leading order in $\alpha_s$ we would expect the difference between
the integrated cross-sections for $a_T$ and $p_T/2$ to tend to a
constant at low $a_T$ whose value is predicted in Eq.~\eqref{eq:pred}.
That this is indeed the case can be seen from Fig.~\ref{fig:lo} where
at large (positive) $L$ the result from MCFM tends to our expectation.
Likewise at NLO accuracy we plot in Fig.~\ref{fig:nlo} the difference
between MCFM and the expectation of Eq.~\eqref{eq:pred} (for the
differential distribution, i.e.~the derivative with respect to $L$,
$D(a_T)-\tilde D(p_T/2)|p_T/2 = a_T$ of Eq.~\eqref{eq:pred}). The result
should tend to a constant at large $L$ and we see evidence for this in
Fig.~\ref{fig:nlo}.

\section{Conclusions}
In this article we have provided a resummation of the $a_T$
distribution to NLL accuracy. We have cross-checked our result by
checking its expansion against MCFM. In forthcoming work we shall aim
to match the resummed result to fixed-order results from MCFM and
include non-perturbative effects in order to have a complete
prediction for comparison to experimental data.

\begin{footnotesize}



%

\end{footnotesize}


\end{document}